\documentclass[reprint,           % Format : preprint, twocolumn
               twocolumn,
               noshowpacs,          % Pacs : showpacs, noshowpacs
               nopreprintnumbers,     % Preprint: preprintnumbers,
               aps,                 % Society: 
               prd,                 % Journal Style : pra, prb, prc, prd, pre,
               letter,             % Size : letter
	       superscriptaddress,  % Affiliation (Title) : groupedaddress,
               nofootinbib,         % Footnote: footinbib, nofootinbib
               tightenlines,        % Remove additional spaces in a line
               floats,floatfix    % Floating pictures and tables
               ]{revtex4}

\usepackage{amsmath, amsfonts, amsthm, amssymb, graphicx, color}
\usepackage{mathtools}
\usepackage{soul}

\newcommand{\be}{\begin{equation}}
\newcommand{\ee}{\end{equation}}

\begin{document}

\title{Higher Order Clockwork Gravity}

\author{Florian Niedermann} 
\email{florian.niedermann@nottingham.ac.uk}
\author{Antonio Padilla} 
\email{antonio.padilla@nottingham.ac.uk}
\author{Paul M. Saffin}
\email{paul.saffin@nottingham.ac.uk}
\affiliation{School of Physics and Astronomy, 
University of Nottingham, Nottingham NG7 2RD, UK}

\date{\today}

\begin{abstract}
  We present a higher order generalisation of the clockwork mechanism
  starting from an underlying non-linear multigravity theory with a
  single scale and nearest neighbour ghost-free interactions.  Without
  introducing any hierarchies in the underlying potential, this admits
  a family of Minkowski vacua around which massless graviton
  fluctuations couple to matter exponentially more weakly than the
  heavy modes. Although multi-diffeomorphisms are broken to the
  diagonal subgroup in our theory, an asymmetric distribution of
  conformal factors in the background vacua translates this diagonal
  symmetry into an asymmetric shift of the graviton gears.  In
  particular we present a TeV scale multigravity model with
  ${\cal O}(10)$ sites that contains a massless mode whose coupling to
  matter is Planckian, and a tower of massive modes starting at a TeV
  mass range and with TeV strength couplings. This suggests a possible
  application to the hierarchy problem as well as a candidate for dark
  matter.

\end{abstract}
\maketitle
The exponentially large hierarchy between the electroweak scale and the Planck scale suggests that new physics could be very close to the scale of current collider experiments (see e.g. \cite{hunt}).  Generically, the Higgs mass  is quadratically unstable against radiative corrections coming from any physics in this large ultra-violet window. If we are to retain the notion of naturalness \cite{naturalness}, any new theory must incorporate a mechanism to ensure cancellation between loops, as in supersymmetry \cite{susy}.  Alternatively, the exponential hierarchy could merely be an illusion, with the fundamental scale lying much closer to the electroweak scale, thanks, say, to large \cite{ADD} or warped extra dimensions \cite{RS}, or the weakness of the string coupling \cite{LSTTeV}. Yet another possibility is that the observed vacuum expectation value of the Higgs is just one of a much larger landscape: when these values can be scanned by the theory in some way, we can invoke anthropic considerations \cite{anth1,anth2}, or employ some sort of cosmological relaxation procedure \cite{relaxion} to pick out the observed value. This list of proposals for addressing or rephrasing the hierarchy problem is far from exhaustive and as yet no experimental evidence in favour of any particular model has been forthcoming (see e.g.  \cite{bsm1,bsm2}).  

 With this in mind it is important to continue to explore new ideas. Recently, the {\it clockwork mechanism} \cite{clockwork0,clockwork} was proposed in order to generate a hierarchy between the fundamental scale in the theory and the effective coupling of the zero mode to external sources at low energies. It was originally applied to axions with a view to explaining the super-Planckian decay constants required by cosmological relaxation models \cite{relaxion}. The idea is to have  a modest number of fields, or {\it gears}, $\pi_i$, whose masses mix with some characteristic strength $q >1$. The structure of the mass terms  are governed by an asymmetrically distributed unbroken subgroup of $U(1)^{N}$ in the fundamental theory. This is nonlinearly realised by the ``pion" fields as gear shifts, $\pi_i \to \pi_i+c/q^i$,  that are equal up to a rescaling with increasing powers of $1/q$.  The result is a zero mode whose overlap with each of the gears also scales asymmetrically, $a_0  \propto \pi_0+\pi_1/q+\ldots  \pi_{N}/q^{N-1}$. By coupling external sources to one end of the clockwork we are able to engineer very little overlap with the zero mode thanks to the high power of $1/q$.  At low energies, this leads to  an exponentially large hierarchy of scales from a theory with a single mass scale, and order one parameters. See also \cite{martin} for similar ideas applied to dark energy.
 
 The clockwork mechanism was later generalised to a much wider class of fields in \cite{giudice}, in particular to linearised gravity, where it was used to explain the hierarchy between the electroweak scale and the Planck scale. These generalisations were criticised in \cite{craig} who argued, amongst other things, that one could not apply the clockwork mechanism to non-abelian theories, including gravity.  The claim rested on the assumption that there is no site (i.e. gear number) dependence in the couplings, as one might expect from a fundamental theory free of large parameters, and made use of elegant group theoretic arguments that forbid an asymmetric distribution in the structure of the unbroken subgroup. At the level of the low energy effective field theory, such site {\it independence} might be viewed as  a model dependent statement, making a concrete assumption about the underlying UV theory  \cite{giudicereply}. If we allow site {\it dependence} in the  couplings, we can again obtain interesting phenomenology but one might worry about the origin of this hierarchy at a fundamental level even though we only ever couple to one external site. The clockwork idea has seen a number of interesting applications,  especially in the context  of dimensionally deconstructed set-ups (see e.g.\cite{infl,wimp,dillon, G1, Belg}).

This is something of a linguistic debate about what is and is not a meaningful clockwork but one that teaches us some valuable lessons~\cite{craig,giudicereply}. It is certainly true that the {\it standard} clockwork cannot be obtained from a discrete theory with a single scale \cite{craig}.  Indeed, as we will see by investigating the corresponding non-linear ghost-free multigravity  set-up~\cite{HR}, to obtain the classic clockwork mass matrix of \cite{clockwork,giudice} in the linearised theory one must introduce hierarchies at the level of the underlying non-linear theory.  As in \cite{giudicereply}, we could simply introduce a dilaton to account for these underlying hierarchies. Here we take a very different approach, generalising the clockwork philosophy to the four  dimensional arrays that govern the metric interactions in the discrete multigravity framework. By focusing on a sparsely populated array with nearest neighbour interactions only, we show how the desired asymmetric decomposition of the zero mode can be obtained from an underlying theory with a single scale and no large parameters. This yields  a low energy effective theory of a massless graviton with exponentially suppressed couplings. More specifically, if we were to take $M \sim$ TeV to be unique across all sites, we can generate a low energy effective theory of gravity with Planckian coupling with ${\cal O}(10)$ sites and no exponentially large parameters.

It remains the case (as it must) that the interactions break the symmetry down to the  symmetric diagonal subgroup of diffeomorphisms \cite{AGS, HR,craig}. Nevertheless, as we will explain  below, we can still obtain an asymmetric distribution in how this acts on the canonically normalised metric perturbations if that distribution is also present in the background value.  It turns out that the form of the zero mode, the massless graviton, is completely fixed by the structure of the underlying vacua, and that this can be rendered asymmetric with only very mild assumptions on the underlying non-linear theory.
 
 Our starting point is the general action for a ghost-free multigravity theory described by \cite{HR,multimetric}
\be
S =  S_\mathrm{K} + S_\mathrm{V} \;,
\ee
where the ``kinetic'' part for the $N$ metric fields $(g_i)_{\mu\nu}$ is 
\be
  \label{eq:2}
S_\mathrm{K} = \sum_{i=0}^{N-1} \frac{M_i^2}{2} \, \int \mathrm{d}^4 x \, \sqrt{-g_i}\,  R[g_i]\,.
\ee
 Here we include possible site dependence in the spectrum of Planck scales, although we emphasize that we have in mind that each $M_i$ is of order a unique underlying scale, $M$.  It is  convenient to express the potential  in  terms of vielbeins, $(E_i)^a_{\mu}$, such that \cite{HR}
 \be
  \label{eq:3}
S_\mathrm{V} = - \sum_{i,j,k,l}  \int  T_{ijkl}  \epsilon_{abcd}  (E_i)^a  \wedge (E_j)^b \wedge (E_k)^c \wedge (E_l)^d,
\ee
where $(E_i)^a \, = \, (E_i)^a_{\mu} \, \mathrm{d}x^\mu$ and $ (g_i)_{\mu\nu} = \eta_{ab}\, (E_i)^a_\mu \, (E_i)^b_\nu$. Here and in the following, the sums run from $0$ to $N-1$ (unless otherwise stated).
The interaction matrix $T_{ijkl}$ is required to be totally
symmetric and is assumed to depend on the unique underlying scale $T_{ijkl} \sim M^4$. The potential part breaks $N$ copies of the diffeomorphism group acting at each site, down to the diagonal subgroup. Working in the vielbein formalism, $N$ copies of local Lorentz invariance are also broken down to their diagonal subgroup by the potential.

The equivalence between the vielbein and an explicit metric formulation is not automatic. Indeed, if we go beyond pairwise interactions and/or allow ``cycles" of interactions between sites e.g $1 \to 2 \to 3 \to 1$, the equivalence is broken because  the field equations  no longer imply a symmetric vielbein condition \cite{HR}. Such structures, in either vielbein or explicit metric formulations generically lead to ghosts \cite{multimetricghosts,cycles,rescue}  (see \cite{agnis,fawad} for recent constructions that evade this rule).  For nearest neighbour interactions only, as we consider here, we have a chain of pairwise interactions  linking each of the sites $0 \to 1 \to \ldots N-2 \to N-1$, rather than a cycle, and this means the vielbein formulation is equivalent to a metric one and the theory is ghost-free.
 
 The theory admits $N$ Minkowski vacua,
$(\bar{g}_i)_{\mu \nu} = c_i^2 \eta_{\mu \nu}$, provided the constants
$c_i$ fulfil
\be
  \label{eq:6}
\sum_{j,k, l}  c_jc_k c_lT_{ijkl} =  0 \,.
\ee
Note that the $c_i$ cannot all be gauged to unity because there is only one diagonal copy of diffeomorphisms left intact by the potential. In this sense their values are physical up to an overall normalisation.  There does exist a pseudo-symmetry that allows us to conformally rescale each metric $ (g_i)_{\mu\nu} \to \lambda_i^{-2}  (g_i)_{\mu\nu}$ at the expense of rescaling the couplings $M_i \to  \lambda_i  M_i$, $T_{ijkl} \to  \lambda_j \lambda_j \lambda_k \lambda_l T_{ijkl}$. Since we want to work in a frame in which all scales in the action correspond to the unique underlying scale $M$, without any large parameters, this pseudo-symmetry is essentially fixed and cannot be used to remove the $c_i$ in our background solution.  

The overall normalisation of the $c_i$ is fixed by the matter Lagrangian. As we will explain below, matter is only allowed to couple to one particular site and  we normalise all of  the conformal factors relative to this site.  This ensures that any mass scales appearing in the  matter Lagrangian correspond to the physical masses for the canonically normalised matter fields propagating  on the background geometry.

We now consider fluctuations  about our vacua 
\be
  (g_i)_{\mu\nu} = c_i^2 \, \eta_{\mu \nu} + \frac{c_i}{ M_i} \, (h_i)_{\mu \nu} \,, \label{eq:fluct2}
\ee
where the normalisation ensures a canonical form of the Fierz-Pauli kinetic term \cite{FP}.  Thanks to the symmetric vielbein condition\footnote{The symmetric vielbein condition follows from the field equations whenever there are pairwise interactions only and  no cycles, as is the case here \cite{HR}},  $\eta_{ab}\, (E_i)^a_{[\mu} \, (E_j)^b_{\nu]}=0$,  we can use the one diagonal copy of local Lorentz invariance so that the vielbein fluctuations  are symmetric and correspond to 
\begin{align}
  \label{eq:7}
\delta_a^\mu \,  (\delta E_i)_\nu^a \, = \frac{1}{2 \, M_i} \, (h_i)_\nu^\mu \;,
\end{align}
where Lorentz indices are raised and lowered with $\eta_{\mu  \nu}$. 
The second variation of the potential then becomes
\be
  \label{eq:8}
\delta_2 S_V \, = \, \int \mathrm{d}^4 x \, \sum_{i,j} \mathcal{M}_{ij} \, \left[ (h_i)_\mu^\mu \, (h_j)_\nu^\nu - (h_i)^\mu_\nu \, (h_j)^\nu_\mu \right]\, ,
\ee
where the mass matrix is given by
\be
  \label{eq:11}
  \mathcal{M}_{ij}  =  \frac{3}{M_i \, M_j} \,\sum_{k,l} \, c_k\,c_l \, T_{ijkl}\;.
\ee
Let us now choose a frame for which $M_i=M$ for all $i$.  We immediately see the presence of the zero mode $(a_0)_{\mu\nu} \propto \sum_j c_j (h_j)_{\mu\nu}$ from equation \eqref{eq:6}.  To obtain the desired asymmetric distribution, we only really require that $c_j/c_{j+1}={\cal O}(1)>1$. However, for simplicity let us suppose that the  $T_{ijkl}$ are such that the $c_j= c \, q^{-j}$ exactly,  for some overall normalisation constant, $c$, as an example of the desired asymmetric distribution in the overlap between the zero mode and the graviton gears. This is a consequence of the diagonal subgroup of diffeomorphisms  applied to fluctuations on an asymmetric distribution of background vacua. To see this we note that the diagonal diffs act on the metric fluctuations  as $\delta (g_i)_{\mu\nu} \to \delta (g_i)_{\mu\nu}+2c_i^2 \partial_{(\mu} \xi_{\nu)}$ where  $\xi_\mu= \eta_{\mu\nu} \xi^\nu$ is site independent.  In terms of the graviton gears this reads as $ (h_i)_{\mu \nu} \to  (h_i)_{\mu \nu} + 2Mc_i \partial_{(\mu} \xi_{\nu)}$, which is analogous to the asymmetric gear shifts familiar to the original clockwork proposal \cite{clockwork}.  It follows that the form of the zero mode is entirely dictated by the conformal factors in the background vacua and the unbroken diagonal  subgroup of diffeomorphisms.  If those conformal factors exhibit the desired asymmetry then the zero mode has a classic clockwork distribution. In \cite{craig} this possibility was not considered as it was assumed that $c_i=1$ for all $i$. Of course, one might expect that an asymmetric  and hierarchical distribution  in the $c_i$ is not possible without introducing dangerously large  hierarchies in the $T_{ijkl}$, although as we will now show, this is not the case. 
 
 To proceed, we recall that we are assuming nearest neighbour interactions only, consistent with the ghost-free assumption. This implies that the interaction matrix $T_{ijkl}=\tau_{(ijkl)}$ where 
 \be
 \tau_{ijkl}=A_{ij }\delta_{jk} \delta_{kl}+B_{ik} \delta_{ij} \delta_{kl}
 \ee
 The first term above forces three identical indices while the second forces two pairs of identical indices.  Both matrices $A_{ij}$ and $B_{ij}$ are of tri-diagonal form and can be expressed as
 \be
 A_{ij}=\lambda_i^{A} \delta_{ij}+\mu_i^{A} \delta_{i,j-1}\theta_{i,N-1} \theta_{j, 0} +\nu_{j}^{A} \delta_{i-1,j} \theta_{i, 0} \theta_{j,N-1} 
 \ee
 where $\theta_{ij}=1-\delta_{ij}=\begin{cases} 0 & i=j \\ 1 & i \neq j \end{cases}$ and a similar expression given for $B_{ij}$ in terms of $\lambda_i^B, \mu_i^B, \nu_i^B$. Basically, the diagonal components are given by $\lambda_i^{A, B}$, the upper diagonal by $ \mu_i^{A, B}$ and the lower diagonal by $\nu_i^{A, B}$.  In terms of $A_{ij}$ and $B_{ij}$, we have a mass matrix \eqref{eq:11} proportional to 
 \begin{multline} \label{vaccon}
 \sum_{k,l} \, c_k\,c_l  T_{ijkl}=\frac14 (A_{ij} c_j^2 +A_{ji}c_i^2)+\frac12 \delta_{ij} c_i \sum_k A_{ki} c_k\\
 +\frac23 B_{(ij)}c_ic_j+\frac13\delta_{ij}\sum_kB_{(ik)}c_k^2
 \end{multline}
 and a vanishing vacuum condition \eqref{eq:6} given by
 \begin{multline}
\sum_{j,k, l} c_jc_k c_lT_{ijkl} = \frac14 \sum_j (A_{ij}c_j^3+3A_{ji}c_j c_i^2) +\sum_jB_{(ij)}c_i c_j^2
 \end{multline}
 Note that the antisymmetric part of $B_{ij}$ drops out which means we could identify $\mu^B_i$ with $\nu_i^B$. In any event, assuming vacua with  $c_i=cq^{-i}$, the vanishing of \eqref{vaccon}  yields a very weak condition of the form
 \begin{multline}
 \lambda_i^A+\lambda_i^B=-\frac14\left(\mu_i^A \frac{\theta_{i, N-1}}{q^3}+3\mu_{i-1}^A q \theta_{i,0}\right)\\
 -\frac14\left(\nu_{i-1}^A \theta_{i,0}q^3+3\nu_{i}^A \frac{ \theta_{i,N-1}}{q}\right)\\
 -\frac12 (\mu_i^B+\nu_i^B)\frac{\theta_{i, N-1}}{q^2}-\frac12 (\mu_{i-1}^B+\nu_{i-1}^B)\theta_{i, 0}q^2
 \end{multline}
 This implies\footnote{This is obvious in the $B$ sector. In the $A$ sector we can see it by assuming $\mu^A_i \sim q, ~\nu^A_i \sim 1/q$.} that in order to obtain the desired asymmetric distribution in background conformal factors,  we only need to tolerate hierarchies in $T_{ijkl}$ at  order  $q^2$.  Furthermore, we note that we have focused on a special case for which $c_j=cq^{-j}$. Detuning this choice of the $\lambda, \mu, \nu$ by order one (in units of $M$) would simply induce a relative correction of order one to the $c_j$, which will generically preserve the desired hierarchy in the conformal factors.  Perturbing the theory to include couplings that lie off the tridiagonal will weaken the efficiency of the resulting ``clockwork". However, recall that that such a deformation would generically introduce new, ghost-like degrees of freedom that are not expected to be radiatively generated below the cut-off. We shall elaborate on this  later. In any event,  the  mass matrix for our chosen parametrisation is given by
 \begin{multline} \label{massmatrix}
  \mathcal{M}_{ij}  =  \frac{3c^2 q^{-2i}}{M^2}\left[ -\delta_{ij} \left(\frac{\theta_{i, N-1}}{q}Z_i^++q\theta_{i,0}Z_i^-\right) \right. \\
  \left.+\delta_{i,j-1}\theta_{i,N-1}Z_i^+ + \delta_{i-1,j} \theta_{i, 0} Z_{i}^- \right]
 \end{multline}
 where 
 \be
 Z_i^+=\frac14\left(\frac{\mu_i^A}{q^2}+\nu_i^A\right)+\frac{1}{3q}(\mu_i^B+\nu_i^B)\\
 \ee
 and $Z_i^-=q^2Z_{i-1}^+$. We now see the difficulty in generating the classic clockwork mass matrix of \cite{clockwork, giudice} in the absence of exponentially large hierarchies in the underlying theory.  In \cite{clockwork, giudice}, the mass matrix for the field fluctuations depends on a single overall scale. This is not the case in \eqref{massmatrix} unless the $\mu_i, ~\nu_i$ are chosen to absorb the exponential pre-factor of $q^{-2i}$. Such a choice would amount to choosing a hierarchy of scales in $T_{ijkl}$.  This result could have been anticipated from the no-go claims of \cite{craig}.  Of course, the presence/absence  of hierarchies in $T_{ijkl}$ is only a meaningful statement up to possible conformal rescalings of the metric. However, we recall that we have chosen to work in a conformal frame in which all the Planck scales $M_i=M$, so there is  no ambiguity in what we are saying here.
 
 Given that the mass matrix for gravitons is an emergent object, not independent of the background, we would argue that there is actually no compelling reason for us to require it to depend on a single scale, as in \cite{clockwork}. Instead we choose to impose the single scale requirement at the level of the fundamental theory, the $T_{ijkl}$,  and ask whether or not the spectrum of fluctuations about consistent vacua give rise to an emergent hierarchy, with a zero mode that is exponentially more weakly coupled to external states than the heavy modes. This is certainly possible with the set-up described in this paper. Our clockwork is really a higher order one governed by the four-point vielbein interactions. Although there are no large parameters in the potential, it admits an exponential distribution of conformal factors in the corresponding vacua.  This in turn yields a graviton zero mode with a classic asymmetric clockwork decomposition in graviton gears. 
 
 Let us now study the phenomenology of the mass eigenstates for the graviton fluctuations that emerge from our single scale theory. To simplify the analysis, let us assume that the $\mu_i, ~\nu_i$ are site independent, in other words, $\mu_i^A =\mu^A$ etc. The mass matrix now takes the simple form 
 \begin{multline}
 {\cal M}_{ij}=\frac{F(q)}{M^2} c^2 q^{-2i} \left[\delta_{ij} \left(\frac{\theta_{i, N-1}}{q^2}+q^2\theta_{i,0}\right) \right.\\
 \left.-\delta_{i, j-1}\frac{\theta_{i, N-1}}{q}- \delta_{i-1, j} q\theta_{i,0} \right]
 \end{multline}
 where
  \be
F(q)=-\frac34\left(\frac{\mu^A}{q}+\nu^Aq\right)-\mu^B-\nu^B\\
 \ee
 As anticipated earlier, we also assume that matter is minimally coupled to  a single site, given by $i=i_*$.  
 \begin{align}
  \label{eq:13}
S_m =  \int \mathrm{d}^4 x \,\sqrt{-{g}_{i_*}} \,
\mathcal{L}_m[{g}_{i_*}] \,.  
\end{align}
Note that coupling the same matter to multiple sites will generically yield a ghost \cite{Yama,dR1,zhou}, although in some special cases its mass may exceed  the scale of strong coupling \cite{dR1,dR2, Noller, Solo, rescue} (see, also, \cite{agnis} for novel constructions that remain ghost-free at higher energies). In any event, our conservative choice is a consistent one and yields an effective interaction between the canonically normalised $i_*$th graviton gear and the energy momentum tensor of the form
\be
 \delta S_m = \int \mathrm{d}^4 x \,\frac{1}{2M} \, (h_{i_*})_{\mu\nu} \, T^{\mu\nu} 
 \ee
 where we have used the fact that $c_{i_*}=1$ and $M_i=M$.  The condition on $c_{i_*}$ follows from the fact that we have fixed the overall normalisation of the $c_i$'s relative to the site to which matter couples.  Since $c_i=cq^{-i}$, this fixes the overall conformal normalisation factor to be $c=q^{i_*}$.

 The mass matrix \eqref{eq:13} 
can be diagonalised by a rotation in field space (suppressing
indices), $h_i \, = \, \sum_j O_{ij} \, a_j$. The orthogonal matrix, $O_{ij}$ has its columns given by the unit mass eigenstates. In particular, the zeroth column is given by the unit zero mode so that $O_{i0}={\cal N}q^{-i}$, where ${\cal N}=\left(\sum_{k=0}^{N-1} q^{-2k}\right)^{-1/2}=\left(\frac{1-q^{-2N}}{1-q^{-2}}\right)^{-1/2}$.  Numerical investigations suggest that the $j$th massive eigenstate generically has $O_{i,j>0} \approx 1$, for some $i$. The corresponding massive eigenvalues are given by 
\be 
m_{j>0}^2 \sim \frac{F(q) q^{2(1+i_*-j)}}{M^2}, 
\ee
where we have again used the fact that $c=q^{i_*}$. These results can be obtained analytically in the large $q$ limit, when the mass matrix approximates as $ {\cal M}_{ij} \approx \frac{F(q)}{M^2} q^{2(1+i_*-i)} \delta_{ij} \theta_{i,0}$.

In terms of the mass eigenstates, the coupling to matter reads
\be
 \delta S_m = \int \mathrm{d}^4 x \left( g_0(a_0)_{\mu\nu} +\sum_{j=1}^{N-1} g_j (a_j)_{\mu\nu} \right) T^{\mu\nu} 
 \ee
 where the zero mode coupling is 
 \be
 g_0=\frac{{\cal N}}{2 M q^{i_*}}=\frac{\left(\frac{1-q^{-2N}}{1-q^{-2}}\right)^{-1/2}} {2 M q^{i_*}}
 \ee
 If we couple matter to the end of the clockwork, at site $i_*=N-1$, then for $q>1$ the zero mode coupling is at an exponentially higher scale than the fundamental scale, $M_{0}^\textrm{eff} \sim Mq^{(N-1)}$. Taking $M \sim$ TeV and $q=4$, we can achieve a Planck scale effective coupling $M_{0}^\textrm{eff} \sim M_{Pl}$ with $N=26$ sites. Recall that the level of hierarchy in $T_{ijkl}$ need not exceed $q^2 \sim 16$ in this case.
 
 Turning to the heavy modes, these couple to matter with strength $g_j =\frac{O_{i_*, j}}{2M}$, which is given by the fundamental scale, $M$.  Taking $\mu^{A, B}, ~\nu^{A, B} \sim M^4$, consistent with our single scale theory, there is a mass gap of order $M^2q^2$ to the spectrum of heavy modes. These are then distributed exponentially, with the heaviest mode having a mass, $M^2 q^{2(N-1)}$. Choosing our parameters as in the previous paragraph, this yields lightest and next to lightest heavy modes whose masses lie beyond the TeV scale, with TeV strength and weaker coupling to the energy-momentum tensor. In principle, this spectrum could include an interesting dark matter candidate (see \cite{DM1,DM2,DM3,DM4} for some work on spin two dark matter). 
 
  Before presenting this as a robust solution to the hierarchy problem, we need to ask whether or not the  structure of the potential is radiatively stable. For example, do loop corrections  generate large next-to-nearest neighbour interactions  that could weaken the efficiency of our higher order clockwork? For the case of matter loops the answer is obviously negative since we took matter to only couple to a single site.  For graviton loops, the question is more subtle and the only possible  statement we can make is to ask what happens far below the cut off (TeV) when we treat this as an EFT (if indeed that is a reasonable thing to do). We anticipate that gauge invariance will  prevent zero mode loops from generating any new  potential interactions. 
  Heavy mode loops could be more dangerous although one might just assume that they decouple at low energies
 since the masses start at a TeV scale. Of course, it is possible that decoupling is subtle, at least if the interactions between the light and heavy modes also diverge as we send the masses to infinity. A thorough investigation of this is obviously going to be very involved, as with any calculation involving graviton loops. Indeed, its scope extends beyond the context of this paper to a more general question regarding the radiative stability of ghost-free multigravity theories. This is because additional  beyond nearest neighbour interactions introduce a  trivertex and/or a cycle in our potential, which would resurrect the Boulware-Deser ghost \cite{Noller,rescue}. This represents new degrees of freedom and in  analogy with higher order curvature corrections generated in a perturbative approach to quantum General Relativity,  we might expect them to have mass scales at or above the cut-off of the theory (see \cite{lav,lav2} for a similar statement in a massive gravity and bigravity context). A  more detailed analysis will be very involved but is clearly  a priority for future work.

Another important feature of our model is the absence of an underlying dilaton, in contrast to the original proposals presented in \cite{giudice, giudicereply}. From a four-dimensional perspective, this allows us to have a fully non-linear multigravity clockwork governed by a single (TeV) scale, representing a completely new approach to the electroweak hierarchy problem. The flip side of this particular structure is that it could prove to be an obstacle in obtaining it from a dimensional deconstruction of a five dimensional model. Of course, the ultimate goal would be to realise this set-up as a  string theory compactification. 
 
 To summarise, we have shown that a consistent single scale  multigravity model can yield a clockwork graviton spectrum where the massless graviton couples to matter exponentially more weakly than the heavy modes.  This is achieved through a higher order generalisation of the standard clockwork mechanism involving nearest neighbour interactions in the ghost-free non-linear theory.  Although multi-diffeomorphisms are broken to the diagonal subgroup by these interactions, this translates into  an asymmetric shift of the graviton gears thanks to an asymmetric distribution of conformal factors in the background vacua.  This has led us to a TeV scale multigravity model with ${\cal O}(10)$ sites that contains a massless mode whose coupling to matter is Planckian, and a tower of massive modes starting at a TeV mass range and with TeV strength matter couplings. However, before presenting this as a complete resolution of the naturalness question, we emphasise the need to compute radiative corrections including those mediated by graviton loops in an effective description below the cut-off. This will be a priority in future investigations.

\newpage
\vspace{1cm}
\begin{acknowledgments}
  This work was supported by a STFC Consolidated Grant. AP is also funded by a Leverhulme Trust
  Research Project Grant. We would like to thank  E. del Nobile, M. Sloth and SY. Zhou for useful discussions. 
\end{acknowledgments}

\end{document}